\documentclass[11pt]{article}
 \usepackage{geometry}                
 \usepackage{graphicx}
 \usepackage{amssymb}
 \usepackage{epstopdf}
 \title{Integrality of instanton numbers and p-adic B-model}
 \author{Maxim Kontsevich, Albert Schwarz, Vadim Vologodsky }
 \date{}                                           
 
\begin{document}
 \maketitle

  {\bf Abstract. }We study integrality of  instanton numbers  (genus zero Gopakumar - Vafa invariants)
  for quintic 
  and other Calabi-Yau manifolds. We start with the analysis of the case when the moduli space of complex structures is one-dimensional;  later we  show that our methods can be used to prove integrality in general case.  We give an expression of instanton numbers in terms of Frobenius map on $p$-adic cohomology;  the proof of integrality is based on this expression.
 \section{Introduction}
 
  The basic example of mirror symmetry was constructed in [1]. In this example one starts with holomorphic curves on the quintic $\mathcal{A}$ given by the equation
  $$ x_1^5+x_2^5+x_3^5+x_4^5+x_5^5+\psi x_1x_2x_3x_4x_5=0$$ in projective space. (In other words, one considers A-model on this quintic). Mirror symmetry relates this A-model to the B-model on $\mathcal{B}$ (on the quintic factorized with respect to the finite symmetry group $(\mathbb{Z}_5)^3).$  Instanton numbers are defined mathematically in terms of Gromov-Witten invariants, i.e. by means of integration over the moduli space of curves. The moduli space is an orbifold, therefore it is not clear that this construction gives integer numbers. The mirror conjecture proved by Givental [2] permits us to express the instanton numbers  in terms of solutions of Picard-Fuchs equations on mirror quinfic $\mathcal{B}$; however, integrality is not clear from this expression. Gopakumar and Vafa [3] introduced BPS invariants that are integer numbers by definition; it should be possible to prove that    instanton numbers can be  considered as a particular case of GV-invariants.  However, such a proof is unknown; moreover, there exists no rigorous  definition of GV-invariants.
   
   The goal of present paper is to prove the integrality of instanton numbers. However, we will be able to check only a weaker statement: the  instanton numbers become integral after
   multiplication by some fixed number . We work in the framework of B-model definition. In Section 2 we consider  the case when the moduli space of deformations of complex structure on a Calabi-Yau threefold is one-dimensional. The  proof can be 
    generalized to the case when the moduli space is multidimensional (Section 3).  
 The considerations  of the paper
 are not rigorous. To make the paper accessible to physicists
 we have  hidden mathematical difficulties
 in the exposition below.   The paper [5] will contain a rigorous  mathematical proof of the results of present paper.
 
 We will use freely the well-known mathematical results  about
 sigma-models  on Calabi-Yau threefolds; see, for example,
 [6] or [7]. We will follow the notations of [7].

     The proof of integrality of instanton numbers is based on an important statement that 
    these numbers can be expressed in terms of arithmetic geometry.  May be, the fact that physical quantities can be studied in terms of number theory is more significant than the proof itself. 
    
    Instanton numbers we consider can be identified with genus 0
 Gopakumar-Vafa (GV) invariants. GV-invariants can be expressed in terms of Gromov-Witten invariants, but their integrality is not clear from this expression. 
 One can give a condition
  of integrality of GV-invariants in terms of Frobenius map generalizing Lemma 2 of present paper. It seems that
  GV-invariants also can be expressed in terms of $p$-adic
  B-model.
 
 The   relation between topological sigma-models and number theory was anticipated long ago. The existence of such a relation is strongly
 supported by the fact that Picard-Fuchs equations that play important role in B-model appear also in Dwork's theory of 
 zeta-functions of manifolds over finite fields.  
 
  Calabi-Yau
 manifolds over finite fields and arithmetic analog of mirror
 conjecture where considered in very interesting papers by
 Candelas, de la Ossa and Rodriguez-Villegas [8], see also [9].  We go in different direction: our main goal is to obtain the information about
 sigma-models over complex numbers using methods of number theory.
 
 Notice that $p$-adic methods were used in [13] to prove integrality of mirror map for quintic.  The idea to use the Frobenius map
on $p$-adic cohomology to prove some integrality statements
related to mirror symmetry appeared in [15].
 
 In arithmetic geometry one can consider 
 Hodge structure on cohomology; this means that one can
 define the main notions of B-model theory in $p$-adic
 framework  [11] [12]. Our computations are based on the fact
 that in the situation we consider one can obtain the information
 about  the conventional sigma-model from analysis of its $p$-adic analog . This is a non-trivial mathematical fact; however,
 in this paper we will skip the justification of this statement
 referring to [5].  
 
 \section{Integrality of instanton numbers: the simplest case}
  
  Instead of working with A-model we  consider mirror B-model.
  
    Our starting point is the well known formula relating the Yukawa coupling $Y$ in canonical coordinates (normalized Yukawa coupling) to   instanton numbers $n_k$:
    $$Y{(q)}=const+\sum_{d=1}^{\infty} n_d d^3\frac{q^d}{1-q^d}$$
    (This formula is valid in the case when the moduli space of complex structures is one-dimensional; 
    for the quintic const=5.) 
    
{\bf    Lemma 1.} 
    
     Let us assume that 
     \begin{equation}
\sum_{d=1}^{\infty} n_d  d^3\frac{q^d}{1-q^d }= \sum_{k=1}^{\infty} m_k q^k    \label{ }
\end{equation}
If the numbers $n_k$ are integers then for every prime number $p$ the difference $m_{kp}-m_k$ is divisible by $p^{3(\alpha +1)}$ where  $\alpha$ is defined as the number of factors equal to $p$ in the prime decomposition of $k$. Conversely, if 
\begin{equation}
p^{3 (\alpha+1)}  | m_{kp}-m_k     \label {} 
\end{equation}
   for every prime $p$ and every $k$, the numbers $n_k$ are integers. 
\vskip .1in   
   To prove the statement we notice that the following expression for $m_k$ in terms of $n_k$ can be derived from (1):
   \begin{equation}
\label{ }
m_k =\sum_{d | k} n_d d^3
\end{equation}

Let us suppose that $k=p^{\alpha}r$ where $r$ is not divisible by $p$. Then 
 \begin{equation}
m_{kp}-m_k=\sum_{s | r} n_{p^{\alpha +1} s}(p^{\alpha +1}s)^3
\label{ }
\end{equation}
(We are summing over all divisors of $kp$ that are not divisors of $k$, i.e. over all  $p^{\alpha +1}s, $ where  $  s | r  $  .)

 We see immediately that $m_{kp}-m_k$ is divisible by $p^{3(\alpha +1)}$. 
 
 To derive integrality of $n_k$ from this property one can use the Moebius inversion formula
 \begin{equation}
\label{m}
 n_k k^3=\sum _{d | k}\mu (d)m_{\frac{k}{d }}
\end{equation}
where $\mu (d) $ stands for Moebius function. Recall, that Moebius function can be defined by means of the following properties $\mu (ab)=\mu (a)\cdot \mu (b)$ if $a$ and $b$ are relatively prime, $\mu (p)=-1$ if $p$ is a prime number, $\mu(p^{\alpha})=0$ if $\alpha >1$. Again we represent $k$ as $p^{\alpha} r$ where $p$ does not divide $r$ and a divisor  $d $ of $ k $ as $p^{\beta}s$ where $s | r$ and $\beta \leqslant \alpha$. Taking into account that $\mu (p^{\beta}s)=0$ if $\beta >1$ we obtain 
\begin{equation}
\label{6}
n_k k^3=\sum _{s | r}\mu (s)m_{\frac{k}{s }}+\sum \mu (sp)m_{\frac{k}{sp }}=\sum _{s | r}\mu (s) (m_{\frac{k}{s}}-m_{\frac {k}{sp}})
\end{equation}
It follows from our assumption that the left-hand side of (5) is divisible by $p^{3\alpha}$, hence $n_k$ does not contain $p$ in the denominator  (in other words, $n_k$ can be considered as an integer $p$-adic number). In the above calculation we assumed that $\alpha \geqslant 1$; the case $\alpha =0$ is trivial. If the condition (2) is satisfied for every prime $p$ we obtain that the numbers $n_k$ are integers. 

{\bf Lemma 2.}

The numbers $n_k$ defined in terms of $Y(q)$ by the formula (1) are integers if and only if for every prime $p$ there exists such a series $\psi (q)=\sum s_kq^k$ having $p$-adic integer coefficients that 
\begin{equation}
\label{7}
Y(q)-Y(q^p)=\delta ^3 \psi(q)
\end{equation}
Here $\delta$ stands for the logarithmic derivative $q\frac {d}{dq}$
\vskip .1in 
It is easy to check that this lemma is a reformulation of Lemma 1. The $k$-th coefficient of the decomposition of $Y(q)-Y(q^p)$ into $q$-series is equal to 
 $$  m_k-m_{\frac{k}{p}}= m_{p^{\alpha}s}- m_{p^{\alpha -1}s}$$
   if $k=p^{\alpha}s$ and $\alpha \geqslant 1$, and to $m_k$ if $k$ is not divisible by $p$. From the other side, the coefficients of $\delta ^3 \psi (q)$ are equal to $k^3s_k$.
   
 Notice, that Lemma 2 can be formulated in terms of the Frobenius map $\varphi$. This map transforms  $q$ into $q^p$; corresponding map $\varphi ^\ast $ on functions of variable $q$ transforms $f(q)$ into $f(q^p)$.  The formula (7) can be rewritten in the form 
   $$Y-\varphi ^\ast Y=\delta^3 \psi . $$
   The above statements show that it is natural to apply $p$-adic methods attempting to prove integrality of instanton numbers. 
   
   Recall that $B$-model is formulated in terms of Hodge filtration on the middle-dimensional cohomology of Calabi-Yau manifold;  one should consider deformation of complex structure of Calabi-Yau manifold and corresponding variations of Hodge filtration. Analogous problems can be considered in $p$-adic setting.
   
   If we work with mirror quintic $\mathcal{B}$ (and more general if the moduli space $\mathcal{M}$ of complex structures on a Calabi -Yau threefold is one-dimensional) then one can find a  local coordinate $q$ in a neghborhood of maximally unipotent boundary point of  $\mathcal{M}$ (canonical coordinate) and a basis $e^0(q), e^1(q), e_1(q), e_0(q)$ in three-dimensional cohomology that satisfy the following conditions:

 1) The basis $e^0, e^1, e_1, e_0$ is symplectic: $<e_0,e^0>=-1, <e_1,e^1>=1$, all other inner products vanish.
    
 2) Gauss-Manin connection acts in the following way:
\begin{equation}
\label{8}
\nabla_{\delta} e^0     =0 ,   
\end{equation}
 \begin{equation}
\label{9}
  \nabla_{\delta}e^1     =e^0 ,        
 \end{equation}
 \begin{equation}
\label{10}
\nabla_{\delta}e_1      =Y(q)e^1,  
 \end{equation}
 \begin{equation}
\label{115}
\nabla_{\delta}e_0      =e_1  
\end{equation}

 Here $\delta$ stands for logarithmic derivative $q\frac{d}{dq}$ and $\nabla_{\delta}$ for corresponding Gauss-Manin covariant derivative 
 
 3) $$ e^0 \in \mathcal{F}^0 \cap \mathcal{W}_0$$
 $$ e^1 \in \mathcal{F}^1\cap \mathcal{W}_2$$
 $$ e_1 \in \mathcal{F}^2\cap \mathcal{W}_{4}$$
 $$ e_0 \in \mathcal{F}^3\cap \mathcal{W}_{6}$$
 
 Here $\mathcal{F}^p$ stands for the Hodge filtration and $\mathcal{W}_k$ for the weight filtration (the covariantly constant filtration associated with monodromy around maximally unipotent boundary point).
 
 4) $Y(q)={\rm  const} + \sum m_kq^k $ is a $q$-series with integer coefficients $m_k$.
 
 We assume that $q=0$ corresponds to maximally unipotent boundary point of the space $\mathcal{M}$ and that we are working in a neighborhood of this point.
 
 The conditions we imposed specify the canonical coordinate $q$
 and the vectors of the basis only up to a constant factor. One can fix the canonical coordinate and the vectors $e^0(q), e^1(q), e_1(q), e_0(q)$ up to a sign requiring that the vectors $e^0(0), e^1(0)$ form a $\mathbb{Z}$-basis of $W_2.$ ( Recall, that 
 the bundle of cohomology groups can be extended to the point
 $q=0$; one denotes by $W_k$ the weight filtration on the fiber over this point. One can talk about $\mathbb{Z}$-basis  because the fiber over $q=0$ is  equipped by an integral structure
that depends on the choice of coordinate on the moduli space.)

 All of the statements above are well known; see, for example, [7], Chapters  5 and 6. 

We considered quintic as a Calabi-Yau complex threefold. However,
it is  possible to consider it  and $\mathcal{M}$ over $\mathbb{Z}$ or over     the ring $\mathbb{Z}_p$ of integer $p$-adic numbers
and to study its cohomology over $\mathbb{Z}_p$. It is well known that the Hodge filtration and weight filtration on cohomology can be defined also in this case [10], [11], [12]. 
It is natural to assume that in $p$-adic  setting all of the statements 1)-4) remain valid. This can be proven under certain conditions, however,  the proof is not simple.  A rigorous proof for general Calabi-Yau
threefolds is given in [5]. (For quintic
one can derive these statements from known integrality of
mirror map [13 ] and integrality of one of periods.) Notice, that the Yukawa coupling in $p$-adic situation remains the same, but the coefficients $m_k$ are considered as $p$-adic integers.  

It is important to emphasize that in our consideration we should assume that the manifold at hand remains non-singular  after
reduction with respect to  prime number $p$ . This requirement
can be violated for finite number of primes. (For example, the
mirror quintic $\mathcal{B}$ becomes singular after reduction 
mod $5.$)
 
 In the $p$-adic theory there exists an additional symmetry: the
 so called Frobenius map. (In the situation we need the Frobenius map was analyzed in [10], [12].) {\footnote {A transparent explanation of the origin of Frobenius map based on the ideas of supergeometry will be given in  [14]. We are using the Frobenius map in canonical coordinates, but one can define this map in any coordinate system.}} Namely, the map $q \to q^p$ of the
 moduli space of quintics into itself can be lifted to a homomorphism Fr of
 cohomology groups of corresponding quintics. Here $q$ is
 considered as a formal parameter or as a $p$-adic integer; cohomology are taken
 with coefficients in $\mathbb{Z}_p$. We will express instanton numbers in terms of this map, assuming that $p>3$.
 
Notice first of all that the Frobenius map Fr preserves the
weight filtration $\mathcal{W}_k$. It does not preserve the Hodge filtration $ \mathcal{F}^p$,
but it has the following property :
\begin{equation}
\label{11}
{\rm Fr}\mathcal{F}^s\subset p^s\mathcal{F}^0.
\end{equation}
Notice that to prove (\ref{11}) one should assume that $p>3$
(the proof is based on the inequality $s<p$).

The Frobenius map is compatible with symplectic structure
on 3-dimensional cohomology; more precisely, 
\begin{equation}
\label{12}
<{\rm Fr} a, {\rm Fr} b >= p^3 {\rm Fr}<a, b>,
\end{equation}
 where $<a, b>$ stands for the inner product
of cohomology classes. 

It is compatible with Gauss- Manin
connection $\nabla$; namely, 
 \begin{equation}
\label{13}
\nabla _{\delta}{\rm Fr}  a= p {\rm Fr} \nabla _{\delta} a
\end{equation}
 
 The matrix of Frobenius map is triangular; this
 follows from the fact that Fr preserves the weight
 filtration.  Using (\ref {13}) one can check that the diagonal elements $\epsilon _i$ of this matrix obey $\delta \epsilon _i =0;$ hence they do not depend on $q$.  ln the same way one can prove that two neighboring diagonal elements are equal up to a factor of $p$; in other words ${\rm Fr}e^0=\epsilon e^0$ and all other diagonal elements have the form $p^i \epsilon.$  From  (\ref {12}) we conclude that $\epsilon =\pm 1.$ In what follows we assume that $\epsilon =1;$ the modifications necessary in the case $\epsilon =-1$ are obvious. 
 
 Taking into account (\ref{11}) and (\ref{12}) we
  can write
 \begin{equation}
\label{14}
{\rm Fr}e^0=e^0,
\end{equation}
\begin{equation}
\label{15}
{\rm Fr}e^1=pe^1+pm_{12}e^0,
\end{equation}
\begin{equation}
\label{16}
{\rm Fr}e_1=p^2e_1+p^2m_{23}e^1+p^2m_{13}e^0,
\end{equation}
 \begin{equation}
\label{17}
{\rm Fr}e_0=p^3e_0+p^3m_{34}e_1+p^3m_{24}e^1+p^3m_{14}e^0
\end{equation}
 where $m_{ij}\in \mathbb{Z}_p[q]$ are $q$-series with integer
 $p$-adic coefficients. (The powers of $p$ in RHS come from
 (\ref{11}).)
 Using (\ref{12}) we can obtain also that 
 $$-m_{34}+ m_{12}=0,  
 -m_{23}m_{34}+m_{24}+m_{13}=0.$$
 Applying Fr to (\ref{9}) and using  (\ref{13}) we obtain
 \begin{equation}
\label{18}
pe^0+p\delta m_{12}e^0=pe^0.
\end{equation}
This means  that  $m_{12}$ is a constant.  One can prove  [5] that
$m_{12}=0$ . We see that
\begin{equation}
\label{19}
m_{34}=m_{12}=0,
m_{24}+m_{13}=0.
\end{equation}

Similarly, from (10) we obtain 
\begin{equation}
\label{20}
Y(q)-Y(q^p)=\delta m_{23},
m_{23}+\delta m_{13}=0
\end{equation}
hence
\begin{equation}
\label{21}
Y(q^p) -Y(q)=\delta ^2 m_{13}.
\end{equation}
>From (11) we see that
\begin{equation}
\label{22}
m_{34}Y+\delta m_{24}=m_{23},
\delta m_{34}=0, m_{24}+\delta m_{14}=m_{13}.
\end{equation}
 We   know that $m_{34}=0$,
 hence
 $m_{23}=\delta m_{24}$.  
  Using the last equation in (\ref{22})  and (\ref{19}) we conclude that
  \begin{equation}
\label{23}
2m_{13}=\delta m_{14}.
\end{equation} 
Combining this equation with (\ref{21}) we obtain

{\bf Lemma 3.}

\begin{equation}
\label{eqno:24}
Y(q^p) -Y(q)=\frac{1}{2}\delta ^3 m_{14}.
\end{equation}.
\vskip .1in 
{\bf Theorem.}

Instanton numbers are $p$-adic integers if $p>3$.
\vskip .1in 

This statement follows immediately from Lemma 2 and Lemma 3.

The equation (\ref{eqno:24}) together with (\ref{m})
leads to the representation of instanton numbers in terms of Frobenius map:
\begin{equation}
\label{a}
n_{p^ar}=\sum_{d|r}\mu(\frac{r}{d})M_{p^ad}\frac{d^3}{r^3}
\end{equation}
where we assume that $r$ is not divisible by $p$ and use the notation $M_k$ for the coefficients of power series expansion of $-\frac{1}{2}m_{14}$:
$$-\frac{1}{2}m_{14}=\sum M_kq^k.$$
Conversely, one can expess the Frobenius map in terms of
instanton numbers using Lemma 3 and (\ref{19})-(\ref{23}). (Notice,
that we are talking about Frobenius map in canonical coordinates.)  One should notice  that we assumed that 
diagonal entries of the matrix of the Frobenius map are positive;
the assumption that they are negative leads to the change of sign 
of all entries of this matrix.  One should mention also that  our considerations do not fix the value of $m_{14}$ at the point $q=0$; it seems, however, that one can prove that this value is equal to zero.

\section{Integrality of instanton numbers: general case}
In the  considerations of Sec 2 we restricted ourselves to the case of
quintic or, more generally, to the case when the moduli space of complex structures is one-dimensional. 

Let us analyze the case
when the dimension  of moduli space $\mathcal{M} $ of complex structures is
equal to $r>1$. 

Under certain conditions one can find a basis in three-dimensional cohomology consisting
of vectors $e_0 \in I^{3,3},  e_1,...,e_r \in I^{2,2},  e^1, ...,e^r \in I^{1,1}, e^0 \in I^{0,0}$ where $I^{p,p}$ stands for
$\mathcal{F}^p\bigcap\mathcal{W}_{2p}$. 
In appropriate coordinate system $q_1, ...,q_r$  on $\mathcal{M}$
(in canonical coordinates) Gauss-Manin connection acts in the
following way:
$$ \nabla _{{\delta _i}}e^0=0,$$
$$\nabla _{{\delta _i}}e^k=\delta_{ik}e_0, \ \ \ k=1,...,r,$$
$$\nabla _{{\delta _i}}e_j=\sum  _kY_{ijk}(q) e^k, \ \ \ j=1,...,r$$
$$\nabla _{{\delta _i}}e_o=e_i.$$
Here $\nabla _{{\delta _j}}$ denotes the covariant derivative that 
corresponds to the logarithmic derivative $\delta _j= q_j\partial / \partial q_j.$

We are working in the neighborhood of  maximally unipotent
boundary point $q=0$  . We assume that the B-model at hand can be obtained as a mirror of A-model.  Then one can say that the so called integrality
conjecture of [6] ( see also [7], Sec 5.2.2. ) is satisfied; this is
sufficient to derive the above representation for Gauss-Manin
connection. As in the case $r=1$ the conditions we imposed
leave some freedom in the choice of canonical coordinates and of the basis. We will require
that  the vectors $e^0(0), e^1(0),...,e^r(0)$ constitute a $\mathbb{Z}$-basis of $W_2$.

The Yukawa couplings $Y_{ijk}(q)$ can be considered as
power series with respect to canonical coordinates $q_1,..., q_r$;
these series have integral coefficients. (The integrality of these 
coefficients will not be used in the proof; it can be derived
from the integrality of Frobenius map and the formula (\ref{*}).)

Again one can prove [5] that the Gauss-Manin connection has
the same form in $p$-adic situation; the Yukawa couplings
should be considered as elements of $\mathbb{Q}_p[q_1,...,q_r]$
in this case. 

The Frobenius map has the form
$${\rm Fr}e^0=e^0,$$
$${\rm Fr}e^k=pe^k+p(m_{12})^ke^0,$$
$${\rm Fr}e_j=p^2e_j+p^2(m_{23})_{jk}e^k+p^2(m_{13})_je^0,$$
 $${\rm Fr}e_0=p^3e_0+p^3(m_{34})^je_j+p^3(m_{24})_ke^k+p^3m_{14}e^0,$$
 where $(m_{34})^j+(m_{12})^j=0,(m_{23})_{jk}(m_{34})^k+(m_{13})_j+(m_{24})_j+0.$
 
 We can repeat the considerations of the case $r=1$  to obtain
 $$(m_{34})^j=(m_{12})^j=0,$$
 $$Y_{ijk}(q^p)-Y_{ijk}(q)=\delta _i (m_{23})_{jk}=\delta _i \delta _j (m_{13})_k,$$
 $$2(m_{13})_j=\delta _j m_{14}.$$
 We come to a conclusion that
 \begin{equation}
\label{*}
Y_{ijk}(q^p)-Y_{ijk}(q)=\delta _i \delta _j \delta _k (\frac{1}{2}m_{14}).
\end{equation}
This equation permits us to prove integrality of instanton numbers in the case at hand.

Recall that the Yukawa couplings can be represented in the form:
\begin{equation}
\label{y}
Y_{ijk}(q)=const +\sum_s n_s\frac{q^s}{1-q^s}s_is_js_k=const+\sum_{s,d|s}\frac{q^s}{d^3}n_{\frac{s}{d}}s_is_js_k
\end{equation}
Here $s$ is a multiindex:
  $s=(s_1,..., s_r)$ and
$q^s=q_1^{s_1}...q_r^{s_r}.$ In the second sum $d$ runs over
all positive integers dividing the integer vector $s$.

The numbers  $n_s$   can be identified with instanton numbers
of the mirror A-model. The formula (\ref{y}) remains correct in $p$-adic setting if we consider $n_s$ as $p$-adic numbers. 

Comparing the above formula with  (\ref{*})
we see that
\begin{equation}
\label{**}
\sum_{d|s}\frac{n_{\frac{p^as}{d}}}{d^3}=M_{p^as}
\end{equation}
where $M_r$ are coefficients of the power expansion of
$-\frac{1}{2}m_{14}$ and the vector $s$ in (\ref{**}) is not divisible
by $p$. 
Using  Moebius inversion
formula we obtain from (\ref{**}) an expression of instanton
numbers $n_s$ in terms of $M_r$; this expression has the form 
\begin{equation}
\label{b}
n_{p^ar}=\sum_{d|r} d^{-3}\mu(d)M_{\frac{p^ar}{d}}
\end{equation}
 where $r$ is  an integer vector that is not divisible by $p$ and the sum runs over
all positive divisors of $r$.

 It is clear from this formula that instanton numbers are $p$-adic integers for $p>3$. (We use the fact that matrix elements of
 Frobenius map, in particular, $m_{14}$ are power series 
 having $p$-adic integers as coefficients.) Notice, that the appearance of negative power of $d$ in (\ref{b}) does not contradict $p$-adic integrality, because $d$ is not divisible by $p$.

{\bf Acknowledgments.} We  are indebted to A. Goncharov, S. Katz and A. Ogus  for useful discussions.

We acknowledge the stimulating atmosphere of IHES where this project was started. 

The second author was partially supported by NSF grant DMS -0505735 .
	
 The third author was partially supported by NSF grant DMS-0401164.
 
{\bf  References}

1. P. Candelas, X. de la Ossa, P. Green and L. Parkes, A pair of 
Calabi-Yau manifolds as an exacltly soluble superconformal field theory, Nuclear Physics, B359(1991) 21

2. A. Givental, A mirror theorem for toric complete intersections,
alg-geom/9701016

3. R. Gopakumar, C. Vafa, M-theory and topological strings	, I,II,
hep-th/9809187, hep-th/9812127																										

4. V. Vologodsky, On the canonical coordinates  (in preparation)

5. M. Kontsevich, A. Schwarz, V. Vologodsky, On integrality
of instanton numbers (in preparation)

6. D. Morisson, Mirror symmetry and rational curves on quintic
threefolds: A guide for mathematicians, alg-geom/9202004;
Compactifications of moduli spaces inspired by mirror symmetry,alg-geom/9304007; Mathematical aspects of mirror symmetry, alg-geom/960921

7. D. Cox, S. Katz, Mirror symmetry and algebraic geometry,
AMS, 1999

8. P. Candelas, X. de la Ossa and F. Rodriguez-Villegas,
Calabi-Yau manifolds over finite fields, I, hep-th/0012233,II,hep-th/0402133 
 
 9. Daqing Wan, Mirror symmetry for zeta functions, alg-geom/0411464
 
 10. B. Mazur, Frobenius and the Hodge filtration (estimates),
 Ann. of Math.,98 (1973) 58-95
 
 11. J. Steenbrink, Limits of Hodge structures, Inv. Math. 31 (1976)

 12.  G. Faltings, Crystalline cohomology and p-adic Galois representations.
Alg. Analysis, Geometry and Number Theory, (1989)

 13. B. Lian, S. T. Yau, Mirror maps, modular relations and hypergeometric series, hep-th/9507151
 
 14. A. Schwarz, I. Shapiro , Supergeometry and arithmetic geometry
 (in preparation)
 
 15. J. Stienstra, Ordinary Calabi-Yau -3- Crystals, alg-geom/212061,
The Ordinary Limit for varieties over $Z[x_1,..,x_r]$, alg-geom/0212067

 \vskip .2in
 IHES, Bures-sur-Yvette, 91440, France
 
 maxim@ihes.fr
 \vskip .2in
 Department of Mathematics, UCDavis, Davis, CA95616, USA
 
 schwarz@math.ucdavis.edu
 \vskip .2in
 Department of Mathematics, University of Chicago,
                                   5734 S. University Avenue,
                                    Chicago, Illinois 60637, USA
                                    
  volgdsky@math.uchicago.edu                                  
 \end{document}